\documentclass[final,5p,times,twocolumn]{elsarticle}
\biboptions{numbers,sort&compress}
\usepackage{amssymb}
\usepackage{lipsum}
\usepackage[utf8]{inputenc} % Only needed if you're not using a modern engine like LuaLaTeX or XeLaTeX
\usepackage{amssymb}
\usepackage{graphicx}
\usepackage{dcolumn}
\usepackage{bm}
\usepackage{braket}
\usepackage{mathrsfs}
\usepackage{booktabs}
\usepackage{multirow}
\usepackage{url}
\usepackage{physics}
\usepackage{subfigure}
\usepackage[colorlinks=true,linkcolor=blue, filecolor=blue,urlcolor=blue,citecolor=blue]{hyperref}
\usepackage{color}
\usepackage{cuted}

\usepackage{overpic}
\makeatletter
\usepackage{epsfig}
\usepackage{amsthm}
\usepackage{tikz}
\usetikzlibrary{quantikz}
\usepackage{comment}
\usepackage{array}
\usepackage{diagbox}
\usepackage{comment}
\usepackage{natbib}
\usepackage{cleveref}
\crefname{section}{Sec.}{Secs.}
\crefname{figure}{Fig.}{Figs.}
\crefname{subsection}{Sec.}{Secs.}
\newcommand{\refe}[1]{Eq.~(\ref{#1})}
\newcommand{\reff}[1]{\cref{#1}}
\newcommand{\refs}[1]{\cref{#1}}

\usepackage{bm}

\usepackage{color}

\makeatother

\journal{Physics Letters B}

\begin{document}
	
	\begin{frontmatter}

\title
{Excited States from ADAPT-VQE convergence path in Many-Body Problems: application to nuclear pairing problem and $H_4$ molecule dissociation }

\author[first]{Jing Zhang}
\ead{jing.zhang@ijclab.in2p3.fr}
\author[first]{Denis Lacroix}
\ead{denis.lacroix@ijclab.in2p3.fr}

\affiliation[first]{organization={Université Paris-Saclay, CNRS/IN2P3, IJCLab},%Department and Organization
	% addressline={}, 
	city={Orsay},
	postcode={91405}, 
	% state={},
	country={France}}

\begin{abstract}
A quantum computing algorithm is proposed to obtain low-lying excited states in many-body interacting systems. The approximate eigenstates are obtained by using a quantum space diagonalization method in a subspace of states selected from the convergence path of the ADAPT-VQE (adaptive derivative-assembled pseudo-Trotter Ansatz variational quantum eigensolver) towards the ground state of the many-body problem. This method is shown to be accurate with only a small overhead in terms of quantum resources 
required to get the ground state. We also show that the quantum algorithm might be used to facilitate the convergence of the ADAPT-VQE method itself. Successful applications of the technique are made to like-particle pairing as well as neutron-proton pairing. Finally, the $H_4$ molecule's dissociation also illustrates the technique, demonstrating its accuracy and versatility. 
\end{abstract}


\end{frontmatter}

\section{Introduction}

Quantum many-body systems originating from atomic nuclei, condensed matter, or molecular systems, with particles ranging from very few to several hundred, provide a perfect playground for current and future applications on quantum computing platforms \cite{mcardle2020,motta2022,ayral2023,ayral2025,fauseweh2024,wu2024}.
The earliest applications of quantum algorithms to nuclear physics emerged around ten years ago \cite{visnak2015,visnak2017}, marked by a first landmark calculation of light-nuclei binding energies via Effective Field Theory \cite{dumitrescu2018} (see also \cite{yeter-aydeniz2020,stetcu2022,siwach2021a,shehab2019,nigro2024}). Quantum computing methods have also been applied to study realistic nuclear shell spectra\cite{yoshida2024,bhoy2024,sarma2023,kiss2022,perez-obiol2023,yang2024,romero2022,costa2024}. Targeting at beyond mean-field behavior, simplified model such as the exactly solvable Lipkin-Meshkov-Glick model also serve as a benchmark for quantum computing\cite{larson2010,zhou2017a,cervia2021,robin2023,baid2024,beaujeault-taudiere2024,gibbs2025,chinnarasu2025,hobday2024,hlatshwayo2024} relevant for nuclei. 
Recent efforts address nuclear dynamics and response, such as radioactivity \cite{bedaque2025}, scattering \cite{wang2024a,wu2024a,sharma2024,zhang2025a}, reaction \cite{turro2023,rethinasamy2024}, and heavy-ion collisions \cite{dejong2021}.

Variational methods based on parametrized wave functions allow for better control of convergence 
on noisy devices, and have become a tool of choice in recent years for approximately describing the ground state of many-body systems \cite{larocca2025,tilly2022,cerezo2021}. 
In the context of nuclear physics, spontaneous breaking of symmetry, like particle number or deformation, plays a key role in designing sufficiently expressive variational ansatz that can be efficiently prepared on a classical \cite{ring1980,blaizot1986,bender2003,sheikh2021} or a quantum computer \cite{lacroix2020}, at the cost of restoring the symmetry in a second step. 
It has been exploited with some success for the like-particle pairing problem using the Variational Quantum Eigensolver (VQE) \cite{ruizguzman2022,ruizguzman2023,ruizguzman2024}, but not as well in neutron-proton pairing problem, which motivates the use of the problem-tailored variant of VQE, ADAPT-VQE \cite{grimsley2019,zhang2024a}.
However, nuclear structure studies 
focus both the ground state and low-lying spectroscopy\cite{ring1980}. 
The main objective of the present work 
is to explore the possibility of extending the ADAPT-VQE 
approach to access a set of low-lying states.

The possibility of extending variational methods to access a set of excited states has already been investigated in recent years, majorly by adapting the penalty function.
For example, some algorithms \cite{mcclean2016,izmaylov2019,zhang2021,zhang2021b,wang2023a} shift the global minimum to the low-lying states by a cost function dependent on the variance of energy, $\langle {(H-\varepsilon)^2}\rangle$, which vanishes only for eigenstate(s) of energy $\varepsilon$.
Some works \cite{nakanishi2019,parrish2019,grimsley2025} exploit the unitary invariance of state orthogonality and optimize a weighted energy sum of mutually orthogonal initial states.  
Certain eigensolver methods \cite{smart2024,higgott2019,jones2019,yordanov2022} energetically penalize overlap with undesired eigenstates.

Another important category of approaches, by restricting the problem to a smaller Hilbert space, is Quantum Subspace Diagonalization (QSD).
The critical difference between various QSD algorithms lies principally in the process of subspace generation.
For a comprehensive review we refer to Ref.~\cite{motta2024} and mention here only few representatives.
The quantum subspace expansion method \cite{mcclean2017} forms the subspace about a reference state  (e.g., the approximate ground state) by excitation operators, 
or in a similar spirit, by powers of the Hamiltonian \cite{seki2021} and its exponential \cite{motta2020,cortes2022,epperly2022,shen2023}.
The possibility of utilizing a fixed (or adaptive) parametrized ansatz circuit has also been explored \cite{huggins2020,hirsbrunner2024}. In the multireference selected quantum Krylov algorithm \cite{stair2020}, orthogonal subspaces are constructed from distinct orthogonal reference states and can reduce the condition number of matrices in the generalized eigenvalue problem. 
In nuclear physics, a method based on an idea analogous to QSD is the generator coordinate method (GCM) \cite{ring1980}, the quantum algorithm version of which was recently developed \cite{zheng2023,zheng2024}, building the subspace adaptively from a pool of UCC excitation operators with low-depth circuits. The hybrid GCM-inspired method proves to be effective in obtaining the energy spectra of the Lipkin model \cite{beaujeault-taudiere2024}.
Finally, a class of techniques gaining momentum is based on sampling methods where measurements of an initially prepared state that approximates the ground state are used to identify the relevant states of the computational basis to expand a subspace\cite{nakagawa2024,barison2025}.

The key idea we explore here is 
whether the states generated during the ADAPT-VQE procedure contain sufficient information about the low-lying states and can be efficiently used within the QSD approach. 
To the best of our knowledge, these aspects have not been investigated so far. In the present work, we first review the pairing problem in \refs{sec:adapt-nppairing} and also discuss the combination of ADAPT-VQE and QSD in \refs{sec:adapt+qsd}. Then in \refs{sec:application}, we attempt to solve a nuclear physics problem involving like-particle pairing or neutron-proton pairing. 
We also propose and successfully test 
a simplified optimization strategy to more efficiently integrate ADAPT-VQE into the QSD framework. 
As further validation of the 
approach, we illustrate its application to the dissociation 
of an $H_4$ molecule into two $H_2$. 

\section{The neutron-proton pairing problem}
\label{sec:adapt-nppairing}

As in Ref. \cite{zhang2024a}, the neutron-proton pairing Hamiltonian studied is:
\begin{eqnarray}
H&=& \sum_{i=1}^{n_B} \varepsilon_{i} \left[ (\nu^\dagger_i \nu_i + \nu^\dagger_{\bar i} \nu_{\bar i}) + 
(\pi^\dagger_i \pi_i + \pi^\dagger_{\bar i} \pi_{\bar i}) \right]
\nonumber \\
&-& %g(1- x) 
\sum_{T_z}  g_V(T_z) {\cal P}^\dagger_{T_z} {\cal P}_{T_z} 
-\sum_{S_z} g_S(S_z) {\cal D}^\dagger_{S_z} {\cal D}_{S_z}.   \label{eq:hamilNP}
\end{eqnarray}
Here $(\nu^\dagger_i, \nu^\dagger_{\bar i})$ [resp. $(\pi^\dagger_i, \pi^\dagger_{\bar i})$] denotes creation operators of neutron [resp. proton] time-reversed states. 
The pair operators 
${\cal P}^\dagger_{T_z}$ and ${\cal D}^\dagger_{S_z}$ are given by 
${\cal P}^\dagger_{T_z} = \sum_{i} P^\dagger_{T_z, i}$  and 
${\cal D}^\dagger_{S_z} = \sum_{i} D^\dagger_{S_z, i}$, with:
\begin{eqnarray}
\left\{ 
\begin{array}{l}
\displaystyle P^\dagger_{1,i} =  \nu^\dagger_{ i} \nu^\dagger_{\bar i} , ~~~~
\displaystyle P^\dagger_{-1,i} = \pi^\dagger_{i} \pi^\dagger_{\bar i} \\
\\
\displaystyle P^\dagger_{0,i} = \frac{1}{\sqrt{2}}  \left[ \nu^\dagger_{i} \pi^\dagger_{\bar i}  +\pi^\dagger_{i} \nu^\dagger_{\bar i}   \right]  \\
\\
\displaystyle D^\dagger_{1,i} =  \nu^\dagger_{i} \pi^\dagger_{ i}, ~~~~
\displaystyle  D^\dagger_{-1,i} =\nu^\dagger_{\bar i} \pi^\dagger_{\bar i} \\
\\
\displaystyle D^\dagger_{0, i} = \frac{1}{\sqrt{2}} \left[ \nu^\dagger_{i} \pi^\dagger_{\bar i}  -\pi^\dagger_{i} \nu^\dagger_{\bar i}   \right] \\
\end{array}
\right. \label{eq:pairoperators}.
\end{eqnarray}
As indicated in the Hamiltonian's one-body term, we assume that each level is four-fold degenerated (called a block) and labeled by $i=1, \cdots, n_B$. 
The single-particle states in block $i$ are ordered as $[n \uparrow, n \downarrow, p \uparrow, p \downarrow]_i$. 
Mapping the pairing problem to a qubit surrogate is achieved by the Jordan-Wigner transformation (see Table~II of Ref. \cite{zhang2024a}).  
$T_z = -1, 0, 1$ (resp. $S_z = -1, 0, 1$) denotes the different isovector
$(S,T) = (0,1)$ [resp. $(S,T) = (1,0)$ isoscalar] pairing channels, associated 
with the coupling strengths $g_V(T_z)$ (resp. $g_S(S_z)$).  
In \refe{eq:hamilNP}, by default, the pairing strengths are taken as $g_V = g_S = g$ for all values of $S_z$ and $T_z$, a case referred to as the ``full'' neutron-proton pairing. 
However, various subsets of pairing channels can be studied by simply setting certain coupling strengths to zero. 
For example, the like-particle pairing corresponds to $g_V(\pm 1)=g$ and $g_V(0)=g_S=0$.
The energies $\varepsilon_i$ are given in units of $\Delta \varepsilon$, and for simplicity, we take $g=\Delta \varepsilon$. 

\section{ADAPT-VQE for subspace generation}
\label{sec:adapt+qsd}

\subsection{Short description of the ADAPT-VQE approach}
\label{sec:adaptsummary}

The ADAPT-VQE \cite{grimsley2019} is a variational method that iteratively approximates the ground state of a Hamiltonian.
This method is already well documented, and we summarize here 
only the main ingredients relevant to the present work. 

The approach starts from a predefined pool of operators 
$\mathcal{G} \equiv\{ G_1, G_2, \ldots, G_\Omega \}$ and an initial ``seed'' state $| \Psi_0 \rangle$. 
At the beginning of the $n$-th iteration, it builds a product wave-function
\begin{eqnarray}
| \Psi_n \rangle &=& \prod_{i=1}^{n} e^{\theta_i G_{\alpha_i}} | \Psi_0 \rangle 
\label{eq:trialadapt}
\end{eqnarray} 
The initial energy gradient at iteration $n$  is 
\begin{eqnarray}
\left. \frac{\partial E_n }{\partial \theta_n}\right|_{\theta_n=0} &=& \langle \Psi_{n-1} |\left[ H, G_{\alpha_n} \right]| \Psi_{n-1} \rangle, \label{eq:gradient}
\end{eqnarray}
and $G_{\alpha_n}$ is the operator in $\mathcal{G}$ that gives the largest initial gradient.
The iterative process is stopped when the energy reduction is below a certain threshold. 

In the original version of the ADAPT-VQE \cite{grimsley2019}, to improve the convergence, it was proposed to readjust all $\theta$ parameters at each step $n$. We call this procedure  {\it full parameter optimization} (FPO) below. 
In a simplified scenario, that we call hereafter {\it single parameter optimization} (SPO), only the parameter $\theta_n$ is optimized at iteration $n$, while the optimal estimation from the previous iteration is reused for parameters $\theta_1$ to $\theta_{n-1}$. 

\begin{figure}[htbp!]
	\centering
 \includegraphics[width=0.8\linewidth]{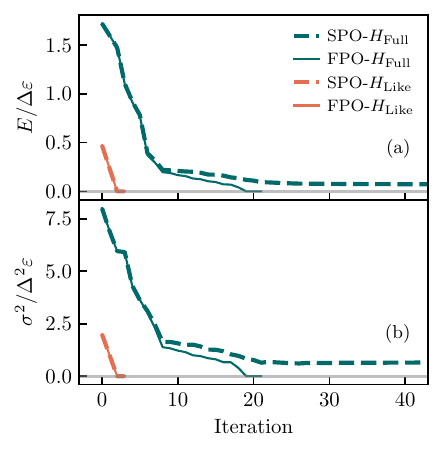}
   \caption{Panel (a): Illustration of the energy, relative to the horizontal gray line of $E_{GS}$, obtained by the ADAPT-VQE for the Hamiltonian (\ref{eq:hamilNP}) with like-particle pairing only interaction ($H_{\rm Like}$) or all pairing channels ($H_{\rm Full}$). 
   The dashed lines show the convergence for the single-optimization technique, while the solid ones show the results of the full optimization. 
   Results are obtained for $3$ protons and $3$ neutrons using the QEB pool operators for three spin-isospin blocks of single particles (corresponding to 12 qubits). 
   The energy $\varepsilon_i$ is $0$, $\Delta \varepsilon$, and $2 \Delta \varepsilon$ for the three spin-isospin blocks. The interaction strength is $g= \Delta \varepsilon$. Panel (b) shows the evolution of the energy fluctuations $\sigma^2_n$, defined in Eq. (\ref{eq:flucn}), during the convergence.
   } 
   \label{fig:twocase-gsonly} 
\end{figure}
The predefined pools $\mathcal{G}$ used in this work are the so-called qubit excitation pool (QEB-pool) \cite{yordanov2021} and the Qubit-pool \cite{tang2021}.  
The QEB pool operators, inspired by operators from the Unitary Coupled Cluster Singles and Doubles (UCCSD) ansatz and simplified for qubit product ansatz \refe{eq:trialadapt}, are given by:
   \begin{equation}
    \left\{
      \begin{array}{lll}
      T^{(1)}_{i j}&=&  \displaystyle\frac{i}{2}(X_{i}Y_{j}-Y_{i}X_{j}),\\[10pt]
      T^{(2)}_{i j k l}&=&  \displaystyle\frac{i}{8}(X_{i}Y_{j}X_{k}X_{l}+Y_{i}X_{j}X_{k}X_{l} \\ 
     &\displaystyle &+ Y_{i}Y_{j}Y_{k}X_{l}+Y_{i}Y_{j}X_{k}Y_{l}-X_{i}X_{j}Y_{k}X_{l} \\
     &\displaystyle &-X_{i}X_{j}X_{k}Y_{l}-Y_{i}X_{j}Y_{k}Y_{l}-X_{i}Y_{j}Y_{k}Y_{l}). 
  \end{array}
  \right.
  \label{eq:qeb}
    \end{equation}
 We have shown in Ref. \cite{zhang2024a} that the QEB pool is, for the pairing problem considered here, the best compromise between (i) the symmetry that might be broken during the optimization, and (ii) the expressivity and convergence performance for the ground state of the Hamiltonian (\ref{eq:hamilNP}). 
 In the discussion of molecule dissociation, we will also use the so-called Qubit-pool \cite{tang2021}. This pool assumes a further simplification by directly considering the Pauli string operators that appear in the QEB-pool, i.e., $\{i X_i Y_j , \; i X_{i}X_{j}X_{k}Y_{l},\; i Y_{i}Y_{j}Y_{k}X_{l}\}.\label{eq:qpool}$
 
 We show in Fig. \ref{fig:twocase-gsonly} a typical illustration of the convergence towards the ground state energy, denoted hereafter by $E_{GS}$, for the particle-like ($H_{\rm Like}$) and full ($H_{\rm Full}$) Hamiltonians. While for the particle-like pairing, performing SPO or FPO does not drastically change the convergence, for the full proton-neutron pairing case, we see that fewer iterations are required in the FPO to accurately reach the ground state energy. In this case, we see that the full convergence towards the ground state is not achieved in SPO.

 \subsection{Hilbert space explored during the convergence}
 
 As a preliminary step, we aim to understand which subspace of the Hilbert space is explored during the descent, and more specifically, if this subspace contains at least some of the low-lying excited states.
 To characterize this subspace, we first consider the energy fluctuation during the convergence, i.e.
 \begin{eqnarray}
\sigma^2_n \equiv\langle H^2\rangle_n  - \langle H\rangle^2_n \label{eq:flucn}  
 \end{eqnarray} 
 using the compact notation $\langle \cdot \rangle_n = \langle n |\cdot | n \rangle$ at iteration $n$. 
 For now, we will use the compact notation $|k\rangle$ for the intermediate state $\ket{\Psi_k}$ with optimal parameters after the $k$-th iteration. 

Such fluctuations are displayed in panel (b) of Fig. \ref{fig:twocase-gsonly} along the ADAPT-VQE path. 
We clearly see that the fluctuations tend progressively to zero when the method converges towards the GS. 
This behavior, together with the decrease of the energy, is a fingerprint of the fact that the ADAPT-VQE approach tends to depopulate gradually the high energy spectrum in favor of the low energy part, leading ultimately to one single state, the GS itself. 
The observation aligns with the Rayleigh-Ritz variational principle: the fidelity of $\ket{n}$ with the ground state is lower-bounded by $\frac{E_{1}- \langle H\rangle_n}{E_{1}-E_{GS}}$ where $E_1$ is the first excitation energy.

\begin{figure}[htbp!]
   \centering
 \includegraphics[width = 0.75\linewidth]{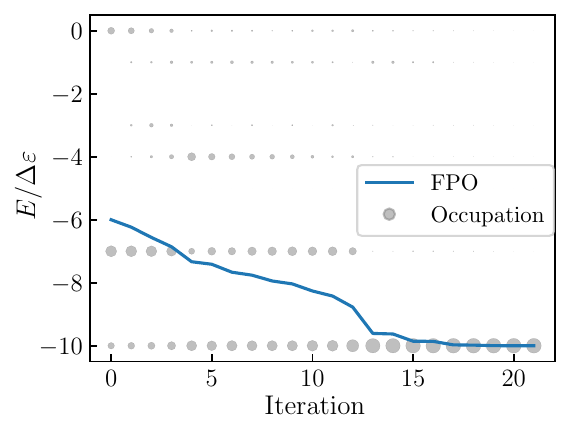}
   \caption{Schematic illustration of the occupation amplitudes $p_n (\alpha)$ as a function of the ADAPT-VQE iteration. This figure is obtained using the QEB pool of operators with the initial state $\ket{000000001111}$ and for three spin-isospin blocks
   with degenerate single particle energies. 
  The blue curve displays the energy $\langle H \rangle_n / \Delta \varepsilon $ as a function of the iteration number $n$.  The gray dots denote the occupation on energy eigenstates at the corresponding eigenenergies. 
  The sizes of the dots are proportional to the ansatz's  $p_n (\alpha)$ values. 
  }
   \label{fig:schematic-dec}
\end{figure}

This aspect is further illustrated in Fig. \ref{fig:schematic-dec}. 
 It schematically shows the occupation amplitudes $p_n (\alpha) = |c_\alpha(n)|^2$ versus iteration step $n$,  
where $| n \rangle = \sum_\alpha c_\alpha(n) | \alpha \rangle$ is expended in the eigenbasis $\{\ket{\alpha}\}$ of the Hamiltonian.  
 During the convergence, we expect that low-energy state amplitudes  
 increase while high-energy states become less populated. 
 At full convergence, the final ADAPT-VQE state matches the ground state with negligible contribution to higher energy states.
An indirect consequence is that the ADAPT-VQE descent may efficiently generate a subspace of many-body states rich in information relevant to low-lying spectroscopy.
In this work, we explore using a Quantum Subspace Diagonalization (QSD) to extract 
these states and probe the low-energy sector of the Hamiltonian.       

\subsection{From ADAPT-VQE to QSD}
\label{sec:gcm}

\begin{figure}[htbp] 
    \centering
    \raisebox{1.5cm}{(a)} 
    \includegraphics[width=0.28\textwidth]{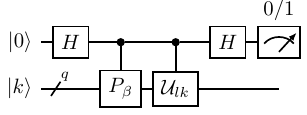}
    \\
    \raisebox{1.5cm}{(b)} 
    \includegraphics[width=0.4\textwidth]{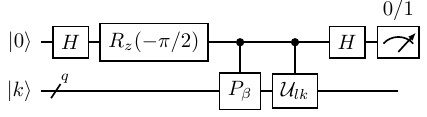}
    \caption{Quantum circuits used to obtain (a) the real and 
    (b) the imaginary part of $\langle l | P_\beta | k \rangle$ for a given Pauli string $P_\beta$. The overlap $\langle l |  k \rangle$ corresponds to the specific case where $P_\beta$ identifies with the identity operator. In both panels, we define ${\cal U}_{lk} \equiv \prod_{i=k+1}^{l}  e^{i \theta_i G_{\alpha_i}}$. 
	Here $H$ denotes a Hadamard gate.}
    \label{fig:expec-p}
\end{figure}

The central idea we would like to explore in the present work is the 
possibility to perform subspace diagonalization of the Hamiltonian in the reduced space of states $\{ |k \rangle\}_{k=0,n_f}$
generated during the ADAPT-VQE descent to the GS. Any state in this subspace 
can be written as: $| \Phi \rangle = \sum_{k=0}^{n_f} c_k | k \rangle$. From the intuitive schematic picture given in Fig. \ref{fig:schematic-dec}, we expect that this subspace automatically contains the relevant information of the low-lying states. Finding approximate eigenstates within the subspace leads to the generalized 
eigenvalue equation:
\begin{eqnarray}
\sum_k H_{lk} c_k = E \sum_k c_k O_{lk}.
\label{eq:geneigen}
\end{eqnarray}
To compute the matrix elements $H_{lk}=\langle l | H | k \rangle$ and $O_{lk} = \langle l | k \rangle$, in practice, 
one can decompose the Hamiltonian $H$ after encoding as a linear 
combination of Pauli strings $P_\beta$, i.e. $H = \sum_\beta g_\beta P_\beta$, which is measurable from the Hadamard test in \reff{fig:expec-p}.
Retaining states along the ADAPT-VQE path with SPO has a practical advantage over FPO: all states can be generated by the same circuits, and one can relate one state $| k \rangle$ to another state $| l \rangle$ through 
$| l \rangle = \prod_{i=k+1}^{l}  e^{ \theta_i G_{\alpha_i}} | k \rangle. \label{eq:connectkl}$ 
Here the $\{ \theta_i \}$ parameters are the optimal parameters decided during the ADAPT-VQE. 

One way of solving \refe{eq:geneigen}  is to first diagonalize the overlap matrix $O$, leading 
to a typically reduced set (size $n_s \le n_f + 1$) of orthonormal states, and then diagonalize the Hamiltonian $H$ in the subspace spanned by the new basis \cite{ring1980}.
The subspace dimension $n_s$ is decided by cutting the eigenspectrum of the overlap matrix, denoted by $\{ \lambda_\alpha\}_{\alpha=0, \cdots, n_f }$, to a certain threshold $\epsilon$, i.e. only eigenstates with $\lambda_\alpha \ge \epsilon$ are retained.

\begin{figure}[tbph!]
	\centering
    \includegraphics[width=0.8\linewidth]{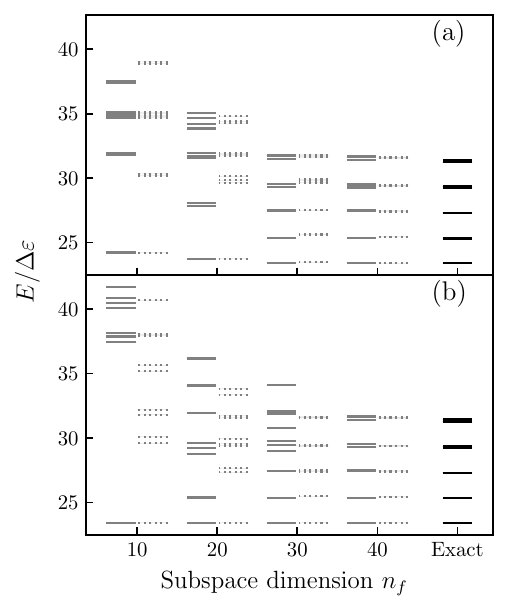}
    \caption{Energy spectra of the like-particle pairing problem obtained by solving the eigenvalue problem shown in Eq. (\ref{eq:geneigen}) using $n_f$  ADAPT-VQE states. 
	The solid and dotted lines are respectively for full optimization (FPO) and the single optimization (SPO). 
	In panel (a), iteration $[0, n_f]$ are retained as input to the QSD, while in panel (b) iterations from $40 - n_{f}$ to $40$ are used.  
	The pair-to-qubit encoding is used \cite{ruizguzman2023} to encode five pairs of the same species (ten particles) populating on ten doubly degenerated equidistant single-particle levels with energies $\epsilon_i = (i-1) \Delta \epsilon$, and we used the coupling strength $g/\Delta \epsilon =1$.
	The matrix elements used in the QSD are obtained using the noiseless ``statevector'' simulator of Qiskit \cite{javadi-abhari2024}, which in practice will be accessed by the circuits shown in Fig. \ref{fig:expec-p}.  
	The exact solution has been obtained by the full CI diagonalization of the Hamiltonian. 
	 Only the lowest $10$ states are compared. 
     }
    \label{fig:gcm-plike}
\end{figure}

\section{Applications}
\label{sec:application}

The QSD method based on the ADAPT-VQE approach is applied below to the pairing problem, either considering only like-particle or the full neutron-proton pairing
Hamiltonian as well as the $H_4$ molecule dissociation process.

\subsection{QSD method applied to the like-particle pairing}

\reff{fig:gcm-plike} illustrates the QSD spectrum for a like-particle pairing problem. 
We also tested how a specific way of expanding the subspace from a set of states can affect the quality of final results. 
In panels (a) and (b), the selection of iterations starts either from the first or the last iteration among the range [0,40]. In each panel, both the SPO and FPO strategies are tested.
As more ADAPT-VQE intermediate states are used, it converges to the exact low-lying spectrum once the subspace dimension $n_f$ exceeds approximately $30$. 
For SPO, the backward selection starting from iteration 40 is also efficient, as shown in panel (b). Note that the subspace dimension here is much smaller than 
that of Full Configuration Interaction calculations in the seniority-zero subspace, which is $C^5_{10}=252$. 

\begin{figure}[htbp!]
  \centering
  \includegraphics[width=0.8\linewidth]{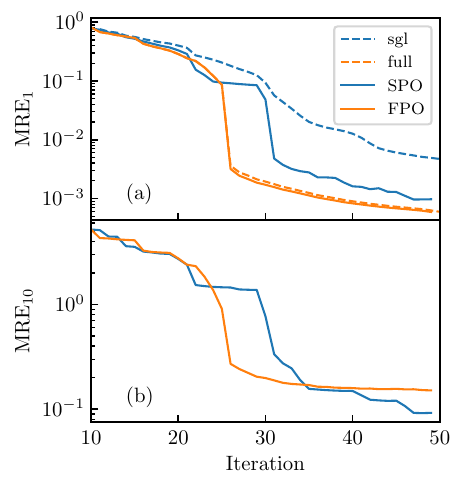}
  \caption{Mean Relative Error obtained from panel (a) of Fig. \ref{fig:gcm-plike} between predicted and exact energy values as a function of the number of iterations for $\Lambda =1$ (a) and $\Lambda=10$ (b). 
  The MRE is shown for the QSD results (solid lines)  as a function of the number of iterations. 
  In panel (a), the MRE obtained after the ADAPT-VQE (sgl and full) before QSD is also shown in a dashed line for comparison. Results from single and full optimization are shown, respectively, in blue and orange. 
  }
  \label{fig:emin}
\end{figure}

To analyze the performance of the ADAPT-VQE + QSD approach, we used the MRE (Mean Relative Error) quantity, ${\rm MRE}_{\Lambda} = \frac{1}{\Lambda \Delta \epsilon} \sum_{i=1}^\Lambda  \left| \overline{E_i}- E^{\rm ex}_i \right|,
\label{eq:MRE}$ where $\{E^{\rm ex}\}_{i=1, \cdots , \Lambda}$ denote the $\Lambda$ lowest exact eigenenergies while $\{\overline{E_i}\}_{i=1, \cdots , \Lambda}$ are the approximate ones.
In the following, we will take either $\Lambda=1$ (GS only) or $\Lambda =10$ (ten first low-lying states). 

The MRE${}_\Lambda$ of different optimization strategies are compared in Fig. \ref{fig:emin}. 
Focusing first on panel (a), we see that the full optimization, where all parameters are optimized, predicts the GS with high precision.
In this case, the FPO result does not descrease MRE${}_1$ much further. 
In contrast, with single optimization, the ground state energy precision improves by an order of magnitude. 
This is an interesting finding that might 
have practical applications. 
On one hand, combining QSD with ADAPT-VQE accelerates convergence towards the global energy minimum.
The resulting ground-state energy is less than or equal to that of the input states, as shown in panel (a) of Fig. \ref{fig:emin}.
On the other hand, full optimization rapidly becomes prohibitive as the number of parameters grows. 
In addition, for the overall accuracy MRE${}_{10}$ in Fig. \ref{fig:emin} (b), the two optimization techniques exhibit alternating performance.
In these contexts, SPO offers a promising alternative to FPO.

\subsection{QSD approach applied to the neutron-proton pairing}
\label{sec:gcm np pairing}

\begin{figure}[htbp!]
	\centering
  \includegraphics[width=.7\linewidth]{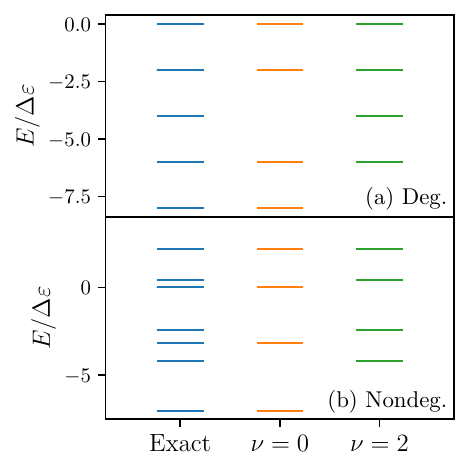}
  \caption{ Illustration of the low-lying spectra obtained for the neutron-proton pairing with two blocks and 4 particles ($2$ neutrons and $2$ protons) using the QSD approach.
  Here, the two panels show spectra for a system with (a) degenerate and (b) non-degenerate $\varepsilon_i$ in \refe{eq:hamilNP}. In the non-degenerate case, the two blocks have a difference of 
  energy denoted by $\Delta \epsilon$ that serves as a scale of energy. 
  The three sets of low-lying states are respectively from left to right: the exact diagonalization (left), the QSD + QEB-ADAPT-VQE initialized with a $\nu=0$ state $| 0000 1111 \rangle$ (middle) or 
  with a $\nu=2$ state $|00011110 \rangle$ (right).
  The interaction strength $g=\Delta \epsilon$ is assumed for all spin-isospin channels in \refe{eq:hamilNP}.}
  \label{fig:qsd-np}
\end{figure}

In this section, we now consider the full neutron-proton pairing problem assuming a uniform two-body 
interaction for all channels. 
Again, the QEB pool operator was used for the ADAPT-VQE that was shown to 
provide a convenient pool for this problem \cite{zhang2024a}. 
In this system, the system size is more limited due to the need for fermion-to-qubit encoding. 
As an illustration, the case of 2 neutrons and 2 protons on two register blocks in Fig. 1 of Ref. \cite{zhang2024a} is considered.  

The generalized eigenvalue problem uses the states generated by
ADAPT-VQE starting from a seniority zero ($\nu=0$) or two ($\nu=2$) initial state in Fig. \ref{fig:qsd-np}.
The threshold takes $\epsilon=10^{-6}$.
When initializing ADAPT-VQE with seniority $\nu=0$, like $\ket{00001111}$ under the encoding in \refs{sec:adapt-nppairing}, the seniority tends to be preserved during the convergence, and as a result, QSD cannot reproduce the spectrum sector of $\nu \neq 0$. 
The missing states, however, can be recovered simply by restarting 
from a state with different seniority, i.e. $\nu=2$ in Fig. \ref{fig:qsd-np}. 
For the low-energy states of all neutron-proton 
pairing systems tested, a good accuracy was achieved.  

More generally, in many physical situations, the Hamiltonian 
is block diagonal due to symmetries.
Provided that quantum hardware is ideal and the predefined pool operators preserve certain symmetry, one can perform a set of ADAPT-VQE calculations, each followed by QSD, to recover eigenstates in different symmetry blocks. 
This aspect will be further illustrated below by the molecule dissociation.

\subsection{Application to the $H_4$ molecule dissociation} 
\label{sec:gcm molecule}

\begin{figure}[htbp]

  \begin{overpic}[width=0.9\linewidth]{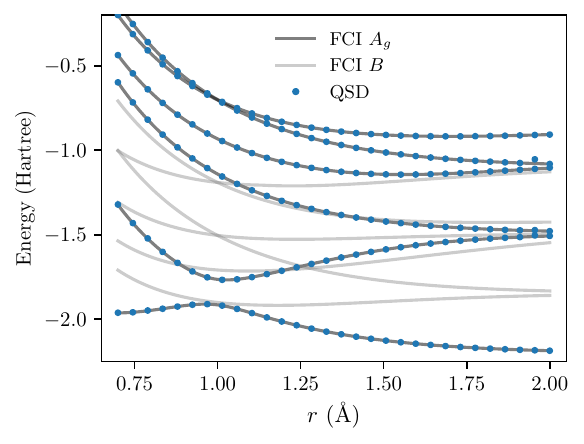}
	\put(73,54){\includegraphics[width=0.21\linewidth]{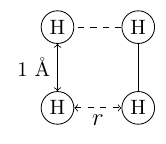}}
  \end{overpic}
  \caption{
  Ground and excited state energies of $H_4$ versus the bond length $r$.
  The planar $H_4$ configuration has adjustable horizontal $H{-}H$ distance $r$ (in \AA) and constant vertical spacing (1 \AA), following Ref. \cite{grimsley2025}.
  Results obtained using ADAPT-VQE followed by QSD are represented by blue dots.  
  FCI results are shown as solid lines for reference. 
  The low-lying states are relevant to several symmetry blocks, 
  and the FCI calculations are made within each symmetry sector. 
  States belonging to $A_g$ point group symmetry, the symmetry of the ground state, are shown in dark gray, 
  while those of other relevant symmetries
  ($B_{1g}$, $B_{2u}$ and $B_{3u}$) are displayed in light gray. 
   The QEB-ADAPT-VQE is applied starting from the state $\ket{00001111}$ and ${\sim}12$ ADAPT-VQE iterations per $r$ value have been used. 
  }
  \label{fig:H4 A1g}
\end{figure}

As the last test of the method proposed, we considered a completely different system, the $H_4$ molecule with the geometry shown in the upper right part of Fig. \ref{fig:H4 A1g}, and its dissociation into two $H_2$ molecules. 
It has become a common benchmark system for quantum algorithms in quantum computational chemistry, particularly for assessing their accuracy and reliability in predicting energy spectral properties \cite{zheng2023,parella-dilme2024,bonet-monroig2023,smart2024,tang2021,grimsley2019}. 

To start, the QEB-ADAPT-VQE is applied for each $r$ value with the initial state $\ket{00001111}$ (two spin-up and two spin-down orbitals occupied). In the noiseless simulation with Qiskit statevector simulator, ADAPT-VQE with FPO properly converges to the ground state at the tested $r$ values. Then QSD is performed with states generated during the ADAPT-VQE procedure.    
The PySCF package provides the Hamiltonian of $H_4$ in the STO-3G basis and the Full Configuration Interaction (FCI) solution \cite{sun2020}. 
In the results from QSD in \reff{fig:H4 A1g}, we observe that the ground state 
and a restricted set of excited states are well reproduced. However, some states 
are completely missed, similar to the neutron-proton case in Fig. \ref{fig:qsd-np}. It might be attributed again to the fact that different low-lying states belong to different symmetry blocks. A full discussion of the relevant symmetries for the $H_4$ planar molecule can be found in Ref. \cite{grimsley2025}. 
In short, the QEB operators selected do not break the initial point group symmetry throughout the optimization. 
Therefore, when the initial state belongs to the $A_g$ symmetry, the QSD can only lead 
to the eigenstates belonging to the $A_g$ subspace, excluding low-lying states within different symmetry sectors.

\begin{figure}[htbp]
  \includegraphics[width=.9\linewidth]{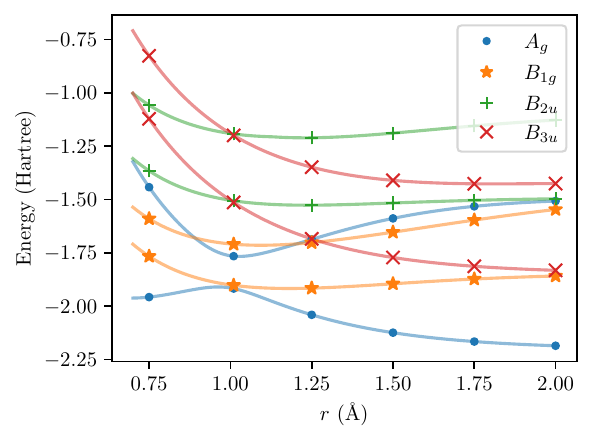}
  \caption{Same as Fig. \ref{fig:H4 A1g} for dissociation energies of $H_4$ but using the state-averaged Qubit-ADAPT-VQE and QSD.
    Using the Qubit pool instead of the QEB pool helps the convergence in states of $B_{1g}$, $B_{2u}$, and $B_{3u}$ symmetries. Consequently, the excited states of those symmetries that are not in Fig. \ref{fig:H4 A1g} can also be found after QSD.
  }
  \label{fig:multi H4 qpool}
\end{figure}

To get the missing states, one might eventually follow the same strategy as illustrated in Fig. \ref{fig:qsd-np}, i.e., one can start ADAPT-VQE from an initial state belonging to a given symmetry sector and obtain the lowest energy state in this sector. 
Then, the QSD is applied to get the set of low-lying states for this symmetry. This strategy was tested 
for the $B_{1g}$, $B_{2u}$, and $B_{3u}$ symmetries. For $B_{1g}$, this procedure indeed produces the expected results. However, for the two other symmetries, 
we encountered the following problem. We see in Fig. \ref{fig:multi H4 qpool}
that the lowest energy states of these two symmetries are degenerate at $r_0=1$ \AA. 
When performing the ADAPT-VQE, we systematically converge to the $B_{2u}$ (resp. 
$B_{3u}$) lowest state below (resp. above) $r_0$. It was never possible, in particular, 
to get the $B_{2u}$ (resp. 
$B_{3u}$) lowest state above (resp. below) $r_0$ and some states of the $H_4$ spectra were systematically missed.

To solve this problem, we finally combined the strategy proposed here with 
the state-averaged method for the ADAPT-VQE that was proposed in Ref. \cite{grimsley2025} to solve 
the ``$B_{2u}$--$B_{3u}$'' level crossing difficulty.
The idea behind the state-averaged strategy is to use the Gross-Oliveira-Kohn variational principle \cite{gross1988} that optimizes towards a set of low-lying states.
This method takes advantage of some VQE variants \cite{nakanishi2019,parrish2019,grimsley2025} to obtain an effect similar to the deflation strategy, where the excited states are searched for in a restricted subspace orthogonal to the known low-lying eigenstates.
The major difference brought by the state-averaged strategy to the 
ADAPT-VQE is to initialize from a set of orthogonal states $\left\{\ket{\psi_1}, \ket{\psi_2}, \dots \ket{\psi_K} \right\}$ where $\braket{\psi_i}{\psi_j} = \delta_{ij}$. A new cost function is 
used during the optimization
\newcommand{\bt}{\bm{\theta}} 
$C(\bt) = \sum_{i=1}^K c_i \langle \psi_i | U^\dag(\bt) H U(\bt) |\psi_i\rangle.$ The parametrized operator $U(\bt)$ is the product operator in \refe{eq:trialadapt}, preparing an ansatz from each $\ket{\psi_i}$. 
It's a unitary transformation, thus preserving the orthogonality of initial states.
According to the GOK variational principle, when the global minimum of $C$ is obtained, $\{U(\bt) \ket{\psi_i}\}$ will locate the lowest $K$ eigenstates.
More discussion on its predictive power and the optimal $c_i$ can be found in \cite{ding2024}. 
Here, we simply take the weight $c_i=1$. 

In general, the convergence becomes more costly when targeting multiple eigenstates simultaneously (e.g., here $K=4$ for $H_4$),
particularly when the states belong to different symmetries.
To alleviate this, the Qubit pool, instead of the QEB pool, is chosen to accelerate convergence towards the state-averaged global minimum, as it allows symmetry breaking in ansatz states.
Fig. \ref{fig:multi H4 qpool} illustrates the final energy levels of the low-lying states of $H_4$. 
Here, $K=4$ initial states are used for the ADAPT-VQE, properly leading to the four lowest states with symmetries $A_g$, $B_{1g}$, $B_{2u}$, and $B_{3u}$. Importantly, the coexistence of the two low-energy states of $B_{2u}$, and $B_{3u}$ is 
solved at the ADAPT-VQE stage when performing the state-averaged strategy as already demonstrated in \cite{grimsley2025}. 
When adding the QSD in a second step, the excited states absent in Fig. \ref{fig:H4 A1g}, where the state-averaged optimization was not used, are now successfully obtained with good accuracy.

We demonstrate that the strategy of using states generated during ADAPT-VQE 
is versatile and compatible with the state-averaged method.
At a given iteration number $ N_{\rm it}$,
the Hamiltonian 
$H$ and overlap $O$ matrix elements are calculated in the $K\times N_{\rm it}$ subspace of states, making the cost $K^2$ times higher. Reduction of the subspace size, which was not explored in this work, is possible, for instance, through improved sampling strategies, such as random sampling \cite{hirsbrunner2024}.

\section{Conclusion}

In a previous study \cite{zhang2024a}, we have demonstrated that the ADAPT-VQE problem is a powerful tool, even in problems exhibiting spontaneous symmetry breaking. We propose here to use the intermediate states generated during the ADAPT-VQE convergence to access not only the improved ground state 
properties but also low-lying states. 
After confirming that the ADAPT-VQE focuses on 
low-lying states during the convergence,  configuration 
mixing is performed with an increasing number of ADAPT-VQE states. The method is illustrated 
for three different types of problems of growing complexity: like-particle pairing, neutron-proton pairing, and molecular dissociation. 
For superfluid systems, provided that separate calculations are carried out for different seniority sectors, low-lying states were obtained without specific difficulty.
Especially, we propose the single-optimization strategy as a promising alternative to the full optimization of ansatz parameters, achieving nearly comparable performance in the ADAPT-VQE plus QSD framework. 
The application to the $H_4$ molecule dissociating 
into two $H_2$, which presents specific level crossing during dissociation, was 
more challenging. We show that the subspace expansion strategy proposed in this work can be generalized 
by using the state-averaged ADAPT-VQE approach with the symmetry-breaking Qubit-pool. 
Such illustration is also relevant for future applications in 
nuclear physics, where potential energy surfaces are often generated and followed by quantum subspace diagonalization, such as the GCM.

\section{Acknowledgments }

The authors thank Y. Beaujeault-Taudi\`ere and S. Baid for the discussions on the ADAPT-VQE at the early stage of this work. This project has received financial support from the CNRS through the AIQI-IN2P3 project.
JZ is funded by the joint doctoral programme of Universit\'e Paris-Saclay and the Chinese Scholarship Council.
This work is part of 
HQI initiative (\href{www.hqi.fr}{www.hqi.fr}) and is supported by France 2030 under the French 
National Research Agency award number ``ANR-22-PNQC-0002''.

%\input{bib.tex}
% cSpell:disable

\end{document}